# Advanced LEACH: A Static Clustering-based Heteroneous Routing Protocol for WSNs

A. Iqbal[1], M. Akbar[1], N. Javaid[1], S. H. Bouk[1], M. Ilahi[1], R. D. Khan[2]

[1]COMSATS Institute of Information Technology, Islamabad, Pakistan.
[2]COMSATS Institute of Information Technology, Wah Cant, Pakistan.

**ABSTRACT**

Wireless Sensors Networks (WSNs) have a big application in heterogeneous networks. In this paper, we propose and evaluate Advanced Low-Energy Adaptive Clustering Hierarchy (Ad-LEACH) which is static clustering based heterogeneous routing protocol. The complete network field is first divided into static clusters and then in each cluster separate Ad-LEACH protocol is applied. Our proposed protocol is inherited from LEACH with a cluster head selection criteria of Distributed Energy-Efficient Clustering (DEEC). This enables Ad-LEACH to cope with the heterogeneous nature of nodes. Due to small static clusters, each node reduces its broadcast message power because it only has to cover a small area. We perform simulations in MATLAB to check the efficiency of Ad-LEACH. The Simulation results show that Ad-LEACH outperforms LEACH and DEEC in energy efficiency as well as throughput.

## I. INTRODUCTION

The technological advancements in the field of hardware design and system on chip has enabled the designers to create minute wireless sensors which performs many DSP operations[1] [10]. These sensors are autonomous objects, which are vastly deployed in random fashion in the required region. Their number can vary from dozens to many thousands according to requirement. They can provide a high quality and fault tolerant network capability [13-18], however their battery life time is limited. This increases the use of Wireless Sensor Networks (WSNs) in variety of applications. From critical medical applications to battlefield surveillance, the demand of WNSs is increasing rapidly.

The legacy centralized algorithms need a comprehensive knowledge of entire network. Whenever, there is an error in transmission or in case any critical node dies, there are a lot of chances that the protocol reaches the bottleneck [2]. The awful and dynamic environment of WSN increases the chances that a node can lose its connection from network. Thus, increases the bottleneck probability of centralized algorithm. On the other hand, distributed algorithms are based on single node which do not require global knowledge of the network. Therefore, a failure of single node does not affect the system critically. This makes the localized algorithm more robust and scalable. Due to the nature of WSN, each node is tightly power constrained. Therefore, the entire network has limited life time.

To extend the network life, an energy efficient routing protocol is required. Many routing approaches have been proposed to extend life time of battery. Clustering is one of them in which nodes are organized into separate clusters. This approach leads to an energy efficient solution. The protocols proposed in [3] and [4] provide energy efficient routing solution by efficiently organizing cluster formation. Data aggregation can be executed with the help of clustering techniques [5] [6]. The large data coming from many nodes is compressed by using data aggregation techniques. Therefore, a small amount of packets are transmitted which contains meaningful information, this leads to an energy efficient solution.

Clustering technique is of great importance in WSN. It is very efficient in data query and broadcast of messages [7] [8]. Either broadcasting a message or collecting useful information within a cluster, Cluster Head (CH) is of enormous help. The clustering algorithms like Low-Energy Adaptive Clustering Hierarchy (LEACH) [6], Hybrid Energy Efficient Distributed (HEED) clustering approach [11] and Power-Efficient Gathering in Sensor Information Systems (PEGASIS) [12] suppose WSN as homogeneous, which is the reason behind their failure in heterogeneous scenarios.

In this paper, we present a new routing scheme for heterogenous networks, which is more efficient than already proposed schemes; LEACH and DEEC. This scheme inherits the characteristics of LEACH and DEEC. We perform extensive simulations in MATLAB to verify the efficiency of our new scheme; Ad-LEACH, in WSNs.

## II. MOTIVATION

LEACH [6] protocol proposes that each node should elect itself as a CH, therefore, first CH selection process is initiated. After the selection of CH, there is a cluster formation phase. After all nodes assign themselves to their respective CH, each CH allocated a time slot to its client node. All nodes use that allocated time slot to communicate with its respective CH.

The stochastic CH selection algorithm is used in LEACH is proposed for homogeneous networks. CH is a node which collects data from all its client nodes, compress it and then transmit it towards Base Station (BS). During this whole time period, CH consumes extra energy. The node selected as CH dies out quickly due to extra energy burden. It is proposed in LEACH that all nodes in network share this load. To divide this additional load uniformly for all nodes, LEACH proposes that every node $s_i = (1,2,3..N)$ becomes CH after $1/p$ rounds [6]. Ideally it is a good solution, however, as the network evolve the energy level of all nodes differ due to different traffic load and their distance from BS. This leads to unfair CH selection and nodes with low energy dies out quickly.

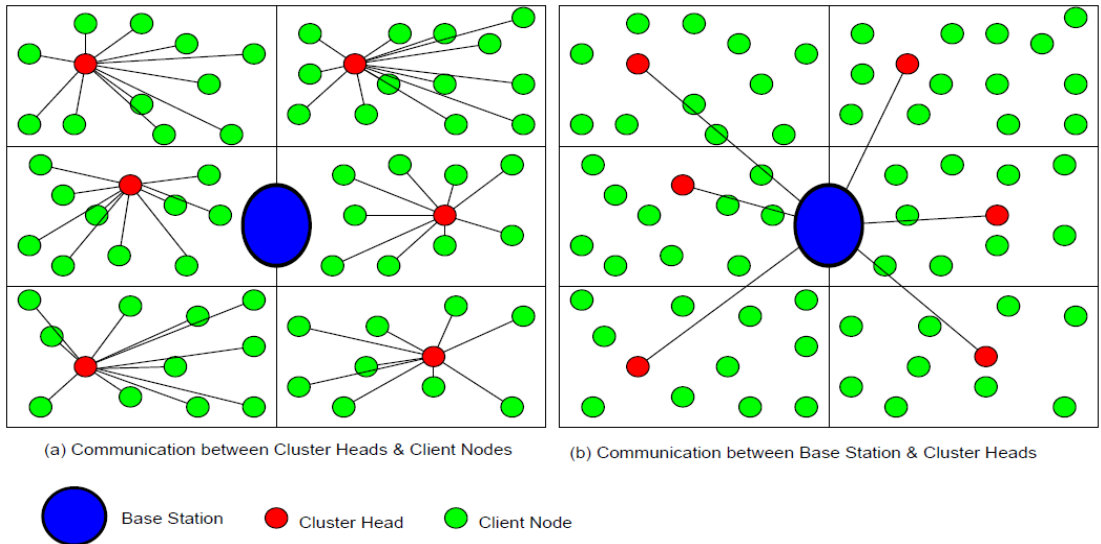

Fig. 1. CH selection criteria of LEACH is given from [6]:

$$T(s) = \begin{cases} \dfrac{p}{1 - p(r \bmod \dfrac{1}{n})} & if\, s_i \in G \\ 0 & otherwise \end{cases} \quad (1)$$

A novel scheme is proposed in this paper which is inherited from LEACH [6] and DEEC [9]. By also taking the legacy static clustering into consideration a protocols is formed which is an energy efficient heterogeneous routing solution.

### III. NETWORK MODEL

In this section, the network model is elaborated. The type of network is heterogeneous. There are $N_{tot}$ numbers of nodes, randomly distributed across $X \times Y$ region. Each cluster contains $N_{cls}$ number of nodes. The distribution of each $N_{cls}$ is random where the value of $cls$ can vary from 1 to q, here we have taken q=4. The main region is further divided into sub regions which are normally referred to as, clusters. The formation of each cluster can be square, rectangular or both according to network design requirement. This type of network deployment is ideal for the scenarios where the physical parameters of geographical area are known, especially for surveillance of high value places. Fig

The nodes are always transmitting data to BS. As shown in Fig. 1 the location of BS is in the center of main region. Fig. 1 (a) shows that all nodes communicate with their respective CH's. Each CH receives data from all of it's client nodes and perform some necessary iterations for compression. Fig. 1 (b) shows that only CH's communication with Base Station. All CH's forward the compressed data to Base Station. As, it is supposed in [10] that all nodes are considered nomadic or stationary within their respective cluster. Therefore, there is no abrupt change in network topology.

There are two types of network on the basis of energy; homogenous and heterogenous. In former type of network, nodes having same energy levels are deployed, whereas, in lateral type, nodes possess different initial energy levels. The heterogeneous network are considered in our work. We divide network into two energy levels of nodes. The nodes with higher energy level are called advanced nodes and the nodes with low energy level are called normal nodes. The percentage of advanced nodes is $m$. Each advanced node posses $a$ times more energy then a normal node. The initial energy of each cluster is equal to $E_{cls}$. Energy computation for clusters, $E_{cls}$, using two levels of energy in [6] is given as:

$$E_{cls} = N_{cls}E_0(1-m) + N_{cls}mE_o(1+a) \quad (2)$$

where, $E_0$ is the initial energy of normal node and total quantity of normal nodes is $(1-m)N_{cls}$. The advanced nodes are $N_{cls}m$ in number and their energy is $E_o(1+a)$. $N_cls$ represents the quantity of nodes present in current cluster.

Therefore, after manipulation we get the energy of each cluster as:

$$E_{cls} = N_{cls}E_0(1+am) \quad (3)$$

The summation gives us the total initial energy of whole area:

$$E_{total} = \sum_{cls=1}^{q} E_{cls} \quad (4)$$

where, $q =$ total number of clusters.

### IV. PROPOSED PROTOCOL

In this section, we discuss our proposed Ad-LEACH in detail. Ad-LEACH is primarily based on static clustering. The location of BS and clustering formation of the entire network is predefined.

### IV.1 CLUSTER FORMATION

In Ad-LEACH, during the establishment of network the whole area is alienated into permanent and static clusters. The shape of clusters can be square or rectangular according to the design requirement and area available. During our simulation we found almost identical results of both rectangular and square shape clusters.
Each cluster contains a separate Ad-LEACH protocol running in parallel to its neighboring clusters. The inspiration in the wake of separating the whole area into small static fields is to reduce complexity and power dissipation. Small portions of clusters are easy to manage rather than one large field of operation. In this way, the nodes also reduce the power level of their broadcast messages because they only have to cover a small portion of area from the main region.

## IV.2 CLUSTERS HEAD (CH) SELECTION

Running a separate Ad-LEACH into all clusters means that each cluster has its own CH. DEEC is proposed in [9], which takes the heterogeneous characteristics of WSNs into consideration. This results in improved scalability and a reduced amount of battery consumption. In order to acquire more definitive solution, we choose CH selection algorithm of DEEC in our Ad-LEACH protocol

The DEEC solution chooses CHs based on their residual energy. Each node requires prerequisite knowledge of network like total energy and network life time. In DEEC, BS broadcasts the total energy of network $E_{total}$ to all nodes. The BS also estimates the value of $R$ which is network lifetime and broadcast it to all nodes. In the start of every new epoch, all nodes calculate the value of $p_i$ using the equation (5) which is taken from [9], as:

broadcasts the total energy of network $E_{total}$ to all nodes. The BS also estimates the value of $R$ which is network lifetime and broadcast it to all nodes. In the start of every new epoch, all nodes calculate the value of $p_i$ using the equation (5) which is taken from [9], as:

$$p_i = \frac{p_{opt} N_{cls}(1+a)E_i(r)}{N_{cls} + \sum_{i=1}^{N_{cls}} a_i \overline{E}(r)} \tag{5}$$

Here $N_{cls}$ is total number of nodes present in current cluster. The value of $\overline{E}(r)$ is calculated in equation(6)[9], as:

$$\overline{E}(r) = \frac{1}{N_{cls}} E_{total}(1 - \frac{r}{R}) \tag{6}$$

Let the energy consumed by network in each round is denoted by $E_{round}$ then the estimated value of $R$ is calculated with equation (7) [9].

$$R = \frac{E_{total}}{E_{round}} \tag{7}$$

Now each node uses the value of $p_i$ and put it in equation number (8) to get the value of $T(s_i)$ [9]. The value of $T(s_i)$ is used by every node to decide if it is CH in current round.

$$T(s_i) = \begin{cases} \frac{p_i}{1 - p_i(r \bmod 1/n)} & if s_i \in G \\ 0 & otherwise \end{cases} \tag{8}$$

After a node is selected as CH, it must keeps its radio receiver turned on so all client nodes to inform the CH about their existence. In order to do that Carrier Sense Multiple Access (CSMA) MAC protocol is used in this phase by all client nodes.

### IV.2.1 PROBABILITY OF HETEROGENEOUS NODES

The equation (9) [9] dictates that $p_{opt}$ is the reference value of $p_i$. In heterogeneous network the reference value of every node differ from each other according to its initial energy value.

$$p_i = p_{opt} \frac{E_i(r)}{\overline{E}(r)} \tag{9}$$

As two level heterogeneous network is considered in this research we will use modified values of $p_{opt}$ as given in equation (10) and (11). [9]

$$p_{adv} = \frac{p_{opt}}{(1+am)} \qquad (10)$$

$$p_{nrm} = \frac{p_{opt}(1+a)}{(1+am)} \qquad (11)$$

This changes the value of $p_i$ and we get equation (12).

$$p_i = \begin{cases} \dfrac{p_{opt} E_i(r)}{(1+am)\overline{E}(r)} & \text{if } s_i = \text{normalnode} \\ \dfrac{p_{opt}(1+a) E_i(r)}{(1+am)\overline{E}(r)} & \text{if } s_i = \text{advancenode} \end{cases} \qquad (12)$$

As we considered two level heterogeneous network in our research we will use modified values of $p_{opt}$ as given in equation (10) and (11).

### IV.2.2 CLIENT SCHEDULING

The CH receives the client information from each node. The CH creates a Time Division Multiple Access (TDMA) schedule for all of its client nodes and broadcast it back to all nodes. this TDMA schedule is used by all client nodes to transmit their data towards CH node.

### IV.2.3 DATA TRANSMISSION

After all client nodes receive their TDMA slot information, the process of data transmission begins. All client nodes only communicate to CH during their assigned time slot. In order to save the energy each client node turned off its radio during unallocated timeslots. The nodes lie near to CH transmit low energy signal and as the distance increases between client node and CH the transmission energy of each node increases. Each client node chooses its own transmission energy level, based on Received Signal Strength (RSS) of the CH advertisement message. When a CH receives data from all of its client nodes, it performs some necessary signal processing techniques on this data to compress it. After compression, this data is transmitted towards BS. During this whole process the radio interface of CH remained turned on, which consumes energy. When CH transmits information towards BS, it is also high energy transmission. This leads to the fact that being a CH puts a lot of energy burden on each node. That is the main reason behind rotating CHs during whole network operation.

### SIMULATION RESULTS

The performance of Ad-LEACH is scrutinized in this section using MATLAB. A heterogeneous WSN containing $n = 100$ nodes is considered in this simulation. The value of $m = 0.1$ which means there are $10\%$ advanced nodes containing a times more energy than the normal nodes. The simulation area is $X = 100m \times Y = 50m$ which creates a rectangular field. All nodes are dispersed around this field randomly. Table. 1 shows the radio parameters used in this simulation.

Table 1: Simulation Parameters

| Parameters | Values |
|---|---|

| $E_{elec}$ | 5nJ/bit |
|---|---|
| $\varepsilon_{fs}$ | $10\,pJ$ /bit/ $m^2$ |
| $\varepsilon_{mp}$ | $0.0013\,pJ$ /bit/ $m^2$ |
| $E_0$ | 0.5J |
| $E_{DA}$ | 5nJ/bit/message |
| $d_0$ | 87.7058 m |
| Message Size | 4000 bits |
| $P_{opt}$ | 0.1 |

In two-level heterogeneous network, LEACH and DEEC is compared with our proposed Ad-LEACH. Fig. 2,3 and 4 show the results using $m = 0.1$ and $a = 0$.

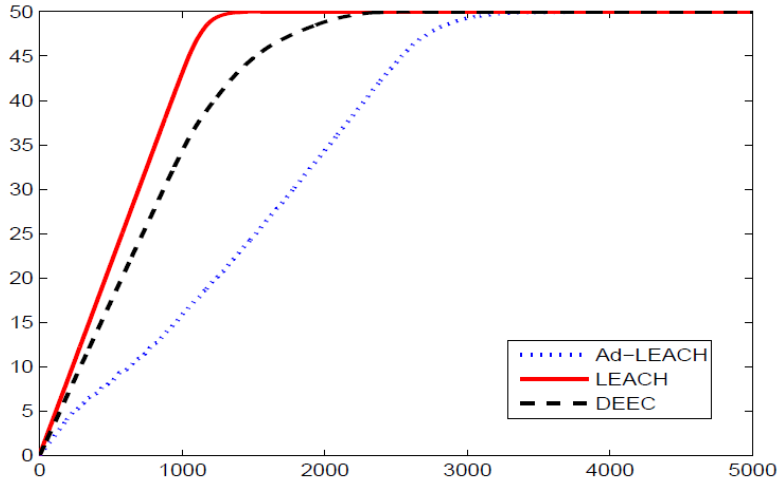

Fig. 2. Energy Consumption in Selected Protocols

Figure 2 shows the result of energy consumption comparison per round between LEACH, DEEC and Ad-LEACH. The results depicts that LEACH consumes more energy and reaches the threshold value of $50\,joules$ between $1000$ and $1500$ rounds. The performance of DEEC protocol is better than LEACH and it reaches the $50\,Joule$ mark in almost $2200$ rounds. Our proposed Ad-LEACH protocol is better than both LEACH and DEEC because of less energy consumption and it reaches the threshold in around $3150$ rounds.

Fig. 3 shows the total number of living nodes per round and Fig. 4 shows total number of dead nodes per round.

The results from these two graphs show that nodes die out in LEACH quickly and in around $1500$ rounds all $100$ nodes die out whereas the first node dies near $1000^{th}$ round. The first node of DEEC dies after $1500$ rounds and the death of last node occurs at $3500$ rounds. The Ad-LEACH shows very promising result here as the death of first node is occurring after $2300$ rounds and the last node dies after $5000$ rounds.

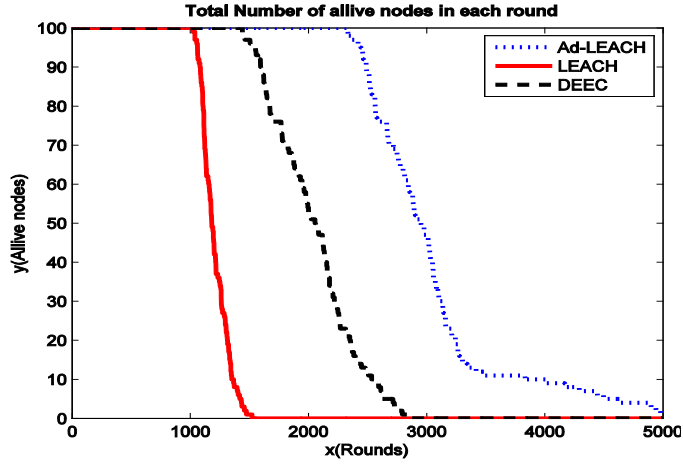

Fig. 3. Comparison of Alive Nodes in Selected Protocols

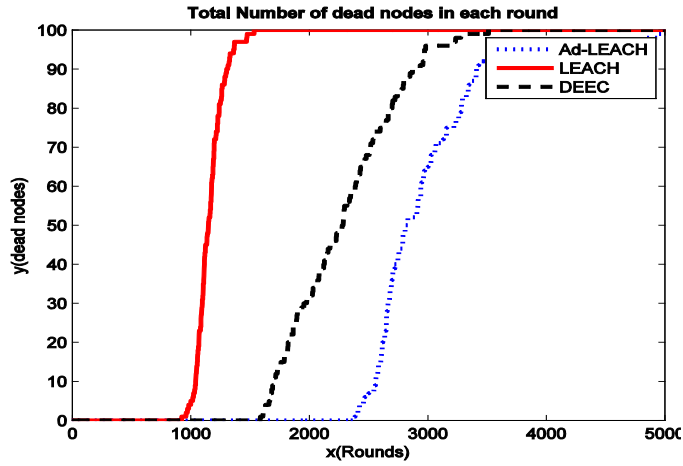

Fig. 4. Comparison of Dead Nodes in Selected Protocols

Stability period of any protocol is when all of its nodes remain alive. After the death of first node till the death of last node, is considered the unstable period. Table (2) compares the unstable and stable regions of LEACH, DEEC and Ad-LEACH.

Table 2: Stable and Unstable Region Comparison of LEACH, DEEC and Ad-LEACH

| Protocol | LEACH | DEEC | Ad-LEACH |
|---|---|---|---|
| Stable region | 1000 rounds | 1500 rounds | 2300 rounds |
| Unstable region | 500 rounds | 2000 rounds | 2700 rounds |
| Network Lifetime | 1500 rounds | 3500 rounds | 5000 rounds |

The throughput comparison is shown in Fig. 5. The figure shows that the throughput of Ad-LEACH is higher than both LEACH and DEEC. In this scenario we changed the values of m and a. The change in these values changed the throughput of Ad-LEACH and DEEC because they both consider the heterogeneous nature of nodes. this change does not affect the performance of LEACH as it is purely designed for homogenous nodes.

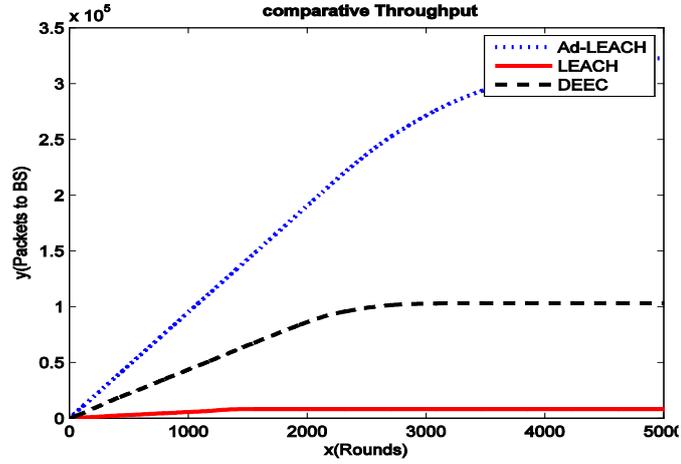

Fig. 5. Throughput Comparison when $m = 0.1$ and $a = 0$

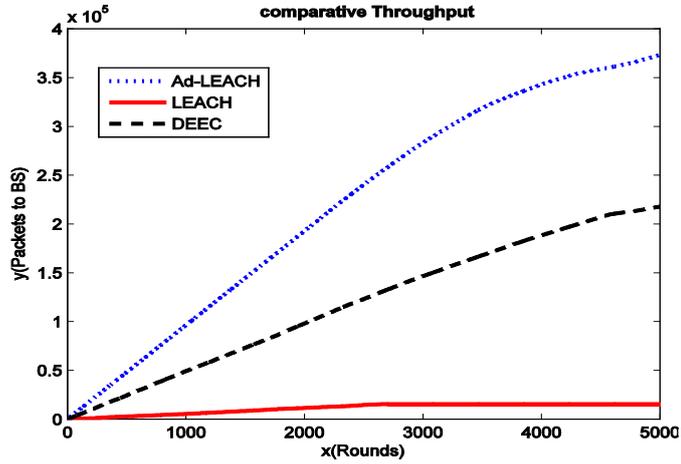

Fig. 6. Throughput Comparison when $m = 0.5$ and $a = 4$

## V. CONCLUSION

In this paper, we propose a new protocol; Ad-LEACH for WSNs. This is an energy efficient routing protocol which is based on legacy static clustering approach. In our proposed scheme, CH selection mechanism is inherited from DEEC whereas, protocol architecture is adopted from LEACH protocol. To validate the performance efficiency of our proposed scheme, simulations are performed in MATLAB. The selected performance metrics are: throughput, energy consumption, number of dead and alive nodes. Simulation results validate the performance efficiency of Ad-LEACH in the case of two level heterogeneous networks, as compared to LEACH and DEEC. Therefore, we conclude that Ad-LEACH is more suitable for heterogenous WSNs.